\documentclass[a4paper,11pt]{article}
\pdfoutput=1 
\usepackage{jcappub} 
\usepackage[T1]{fontenc} 
\usepackage{xcolor}
\definecolor{darkblue}{rgb}{0.0, 0.0, 0.55}
\definecolor{orange}{cmyk}{0,0.5,1,0}
\definecolor{airforceblue}{rgb}{0.36, 0.54, 0.66}

\title{\boldmath Foreground model recognition through Neural Networks for CMB $B$-mode observations}

\author[a,b,c]{F. Farsian,\note{Corresponding author.}}
\author[a,b,c]{N. Krachmalnicoff,}
\author[a,b,c]{C. Baccigalupi}

\affiliation[a]{Scuola Internazionale Superiore di Studi Avanzati (SISSA),\\, Italy}
\affiliation[b]{Istituto Nazionale di Fisica Nucleare (INFN),\\, Italy}
\affiliation[c]{Institute for Fundamental Physics of the Universe (IFPU), \\, Italy}

\emailAdd{ffarsian@sissa.it}
\emailAdd{nkrach@sissa.it}
\emailAdd{bacci@sissa.it}

\abstract{In this work we present a Neural Network (NN) algorithm for the identification of the appropriate parametrization of diffuse polarized Galactic emissions in the context of Cosmic Microwave Background (CMB) $B$-mode multi-frequency observations. In particular, we focus our analysis on the low frequency polarized foregrounds represented by Galactic Synchrotron and Anomalous Microwave Emission (AME). We have implemented and tested our approach on a set of simulated maps corresponding to the frequency coverage and sensitivity represented by future satellite and low frequency ground based probes. The NN efficiency in recognizing the underlying foreground model in different sky regions reaches an accuracy above $90\%$, while the same information using a standard $\chi^{2}$ approach following parametric component separation corresponds to about 70\%. 
Our results indicate a significant improvement when NN-based algorithms are applied to foreground model recognition in CMB $B$-mode observations, and stimulate the design and exploitation of dedicated procedures to this purpose.}

\begin{document}
\maketitle
\flushbottom

\section{Introduction}
\label{sec:intro}
The Cosmic Microwave Background (CMB) and its polarization is a most important probe for Cosmology. The CMB is partially linearly polarized due to Thomson scattering at the epoch of recombination \citep{Hu:1997hv}, and its polarization state can be described by the standard Stokes parameters, $Q$ and $U$, which are coordinate-dependent \citep{Seljak:1996gy}. The CMB polarization pattern can be also decomposed into an alternative base, the $B$ and $E$-modes, with odd and even behavior with respect to parity transformation, respectively \citep{ Kamionkowski:1996ks, Zaldarriaga:2001st}. Unlike $Q$ and $U$ parameters, $B$ and $E$-modes are coordinate-independent on the sphere.\par

Primordial Gravitational Waves (GWs) produced by the  Inflationary era in early Universe are sources of the CMB $B$-mode anisotropies, and represent the main observational target of ongoing and future CMB probes \citep[see][ and references therein]{Campeti:2019ylm}. A second relevant and non-primordial source of $B$-mode anisotropies is represented by the gravitational lensing of CMB photons by forming large scale structure (LSS) \cite{Lewis:2006fu}. The CMB lensing signal is fundamental for investigating the dark cosmological components of the Universe through LSS.\par

The GW contribution to $B$-modes, parametrized by its amplitude relative to primordial scalar perturbations, the tensor-to-scalar ratio, $r$, induces anisotropies at the degree and super-degree scale. The lensing signal dominates the $B$-mode spectrum at the arcminute angular scale \citep{Planck_lensing}.\par

The CMB field is known to possess a Gaussian distribution of anisotropies \citep{Planck_non_gaussianity}, and is characterized primarily through its
angular power spectra. They have been reconstructed with great accuracy over the full sky, for the total intensity ($T$) and $E$-mode polarization, by the Wilkinson Microwave Anisotropy Probe (WMAP) \citep{wmap} and Planck \citep{planckI} satellites. An intense and global effort is currently ongoing towards the measurement of the $B$-mode polarization. Lensing $B$-modes have been detected for the first time  by the South Pole Telescope \citep[SPTpol, see][ and references therein]{spt} through cross-correlation, and directly by POLARBEAR \citep{polarbear_b_modes}. Moreover, they have been observed by the Planck satellite \citep{planck_lens}, the Background Imaging of Cosmic Extragalactic Polarization 2 (BICEP2) \citep{bicep/keck}, the Atacama Cosmology Telescope (ACT) \citep{act}. On the other hand, only upper limits exist so far for the amplitude of the cosmological GWs, corresponding to $r < 0.06$ (at 95\% confidence level) \citep{bkp}. \par

In the last few years, it has become clear that one of the greatest challenges for the detection of primordial $B$-modes is represented by the control and removal of the diffuse emission from our own Galaxy. As a matter of fact, Galactic polarized radiation has an amplitude larger than the cosmological signal on the degree and super-degree scales, at all frequencies and in all the sky regions \citep[see][ and references therein]{planck_component, Krach16, Krachmalnicoff2018}. In order to face this challenge and be able to extract a clean cosmological signal, future CMB probes are characterized by a multi-frequency coverage, with very high sensitivity detectors in all the frequency channels. Along this line, several observatories are currently being built. In particular, The Simons Array \citep[SA, see][]{sa} is being deployed, and the Simons Observatory \citep[SO, see][]{so} will start operations in the early years of this decade. On the longer term, the Stage-IV network of ground-based observatories \citep[CMB-S4, see][]{cmbs4}, along with the Light satellite for the study of $B$-mode polarization and Inflation from cosmic microwave background Radiation Detection (LiteBIRD, \citep{litebird}), are designed to reach an accuracy, after foreground subtraction, corresponding to the capability of detecting a $B$-mode signal with $r$ as low as $10^{-3}$ with a high confidence level.\par

The set of algorithms dedicated to the removal of diffuse foregrounds from the CMB signal is known as component separation, and consists of combining multi-frequency observations in order to reconstruct clean maps of the CMB as well as each foreground emission. In particular, typical methods for component separation are based on parametric fitting of the multi-frequency maps, where the parameters are represented by the amplitude and frequency scaling of the different foreground components \citep{composep, Thorne:2019mrd}. Therefore, a crucial aspect, which constitutes the focus of the present work, is represented by the need of an accurate modeling of the foreground emissions and how the relevant parametrization might vary across the sky, as it is clearly shown in recent and comprehensive analyses concerning proposals of future satellite missions \citep{Core}. An incorrect or inaccurate modeling of Galactic emissions could indeed lead to high residuals in the final CMB maps, preventing the measurement of the faint $B$-mode cosmological signal \citep{cmbs4}.\par

This issue can be thought as a model recognition problem, which represents one of the most important applications of Artificial Intelligence (AI). Neural Networks (NNs) and Machine Learning (ML) in general, as a subset of AI, can be very useful in Cosmology and specifically in the CMB field. In particular, NNs are non-linear mathematical tools characterized by many parameters which are able to model different problems with high complexity. For this reason, they are widely used in science. In the recent years, several works include applications in this direction, ranging from estimating cosmological parameters from dark matter \citep{ravanbakhsh}, to real-time multimessenger astronomy for the detection of the GW  signal from black hole merger \citep{george} and weak lensing reconstruction via deep learning \citep{gupta}. Recent applications, specific to CMB, include: foreground removal from CMB temperature maps using an MLP neural network \citep{tempNN}, convolutional neural networks for cosmic string detection in CMB temperature map \citep{cosmicstring}, lensing reconstruction \citep{deepCMB} and convolutional NNs on the sphere \citep{Krachmalnicoff_2019}. \par

In this paper, we present a new NN application concerning the classification of the appropriate foreground model across the sky, identifying the physical parametrization which describes better a multi-frequency dataset in the different sky regions. This classification has to be seen as a pre-processing to the component separation phase, in order to instruct the latter with the proper functions to be exploited for the fitting. As a case study, in terms of frequency coverage, angular resolution, and sensitivity, we have considered the specifications of the complete frequency coverage of the LiteBIRD satellite \citep{litebird} and the low frequencies channels of the $Q$ and $U$ Joint Observatory in Tenerife \citep[QUIJOTE, see][]{quijote}.
For testing our NN model, we have focused on the analysis on the diffuse Galactic emissions which dominate the low frequency range, i.e. 70 GHz or less, in the CMB $B$-mode observations. Our goal is to study if a pre-processing model recognition phase is possible, and with which efficiency and accuracy.

This paper is organized as follows: in Section \ref{sec:foreground} we review the spectral behavior and parametric models of diffuse Galactic foregrounds which are important for polarization. in Section \ref{sec:NNconcepts} we describe how NNs work, emphasizing the aspects which are particularly relevant for this work, and define the architecture we have implemented and adopted. In Section \ref{sec:res} we study the performance and accuracy of the NN in distinguishing different foreground models distributed differently across the sky. In Section \ref{sec:comparison} we compare the information provided by the NN with the one from a standard approach based on the $\chi^{2}$ statistics using parametric foreground removal. Finally, in Section \ref{sec:conclusion} we bring up the discussion and conclusions.

\section{Foreground Parametrization and Simulations}
\label{sec:foreground}

The main physical mechanisms which are responsible for the emission of linearly polarized light in our own Galaxy are represented by Galactic synchrotron and thermal dust. As we describe in the following, in our analysis, we have also taken into account a third possible source of Galactic polarized radiation, generated by the spin of dust grains, known as Anomalous Microwave Emission (AME), see \citep{planck_component} and reference therein. In the following sections we summarize the main properties of these emissions, and how we simulate them in this work.


\subsection{Synchrotron emission}
\label{sec:synch}
The synchrotron radiation is generated by cosmic-ray electrons accelerating in the Galactic magnetic field. It dominates over the CMB at frequencies $\lesssim$ 70 GHz and possesses a steep Spectral Energy Distribution (SED) due to the corresponding energy distribution of electrons. At first order, the synchrotron SED can be parametrized as a simple power-law in brightness temperature. Nonetheless, the energy distribution of electrons may be responsible for a curvature in the SED, which departs from a pure power-law. The general model for synchrotron emission can be written as: 
\begin{equation}
\label{S3}
T(\hat{n}, \nu) = A_{s}(\hat{n})\left(\frac{\nu}{\nu_0}\right)^{\beta_s(\hat{n}) + C(\hat{n})\log{(\nu/\nu_0)}}\,,
\end{equation}
where $A_s$ is synchrotron amplitude at the pivot frequency $\nu_0$, $\beta_s$ is the synchrotron spectral index, and $C$ represents the SED curvature. In general, all quantities are functions of the sky direction $\hat{n}$. The synchrotron spectral index has a typical value $\beta_s\approx-3$ , with a variation between -2.98 and -3.12 in the sky, on the degree scale \citep{Fuskeland:2014eoa}. In another recent work, the synchrotron spectral index variation has been found to be in the range between -2.5 and -4.4, with a mean value of $\beta_s\simeq -3.2$;  this has been obtained by considering low frequency channels from 2.3 to 33 GHz, combining radio observations by the S-band Polarization All Sky Survey (S-PASS, see \citep{Spass}), WMAP and Planck data \citep{Krachmalnicoff2018}.

Non-zero curvature is suggested by cosmic ray energy spectrum at frequencies above 23 GHz in total intensity, resulting in $C = −0.052 \pm 0.005$ \citep{kogut12}. Krachmalnicoff et al. \citep{Krachmalnicoff2018} have derived an upper limit to the curvature value in polarization: the reported value is between 0.07 and 0.14 depending on the considered sky region and angular scales. \citep{Krachmalnicoff2018}. 

\subsection{Thermal Dust Emission}
\label{sec:dust}
Polarized thermal dust emission \citep[see][and references therein]{Planck_dust} comes from interstellar dust grains which tend to align perpendicularly to the Galactic magnetic field, therefore emitting partially linearly polarized radiation. Dust grains are heated by starlight, and possess a modified black body SED, known as the grey body, with a temperature $T_{d}$ with values around $20$ K and varying across the sky. The SED is also described by a multiplicative emissivity correction $\nu^{\beta_d}$, which determines the deviation from a pure black body, with $\beta_d$ assuming values around $1.6$ and a variation between $1.53$ and $1.67$ across the sky. Thermal dust emission dominates the polarized sky radiation at frequencies $\gtrsim$ 70 GHz \citep[see][and references therein]{planck_component}. The analytic form of the brightness emission of the  SED can be written as: 
\begin{equation}
\label{D1}
    T(\hat{n},\nu) = A_{d}(\hat{n})\left(\frac{\nu}{\nu_0}\right)^{\beta_d(\hat{n})}B(\nu, T_d(\hat{n}))\,,
\end{equation}
where $A_{d}$ defines the dust amplitude varying across the sky at the pivot frequency $\nu_{0}$, and $B$ represents the standard black body spectrum at the temperature $T_d$ and frequency $\nu$ \citep{Planck_dust}. 

\subsection{Anomalous Microwave Emission}
\label{sec:ame}
In total intensity, the AME has been observed by the QUIJOTE telescope and the Planck experiment in the frequency range $\approx$ 10-60 GHz \citep{dickinson2018}. Possible explanations of this emission are represented by spinning dust grains, which rotate at GHz frequencies and emit electric dipole radiation if they possess an electric dipole moment \citep{drain98}, or magnetized dust grains and free-floating ferromagnetic material \citep{drain99}.\par

The AME SED is expected to possess a bell shape, characterized by a peak at around 30 GHz. If the AME is polarized, its polarization fraction must be very small, at the level of a per cent \citep{dickinson2018}. The QUIJOTE, in \citep{quijote_ame}, for example, has constrained the AME polarization to be < 2.8\% with 95\% confidence level in the Perseus molecular complex. In another paper \citep{Genova-Santos:2016bfr} QUIJOTE data allow to put an upper limit to the AME polarization fraction corresponding to 0.39\%, which becomes 0.22\% when combining with WMAP data. Note that the aforementioned limits are related to specific regions and cannot be extended to the whole sky. Remazeilles et al. \citep{Remazeilles:2015hpa} have shown that the AME with 1\% polarization fraction can bias the extracted \textit{r} value, particularly for satellite missions.\par

The model for the AME SED is based on Ali-Haimoud et al. \citep{Ali_2008}. The spinning dust grains with angular velocity $\omega$ and electric dipole moment $\mu$ can radiate as follow:
\begin{equation}
    P = \frac{2}{3}\frac{\mu_{\perp}^2 \omega^4}{c^3}\,,
\end{equation}
where $P$ is the radiation power and $\mu_{\perp}$ is the perpendicular component of $\mu$ to $\omega$. This power is emitted at the frequency $\nu = \omega/2\pi$. The emissivity of electric dipole radiation per Hydrogen ($H$) atom can be calculated through: 
\begin{equation}
    \frac{I_{\nu}}{n_H} = \frac{1}{d\pi}\int_{a_{min}}^{a_{max}}da \frac{1}{n_H}\frac{dn_{gr}}{da}4\pi\omega^2 f_a(\omega)2\pi \frac{2}{3}\frac{\mu_{a\perp}^2 \omega^4}{c^3}\, 
\end{equation}
where $\omega = 2\pi \nu$. The term $\frac{1}{n_H}\frac{dn_{gr}}{da}$ determines the grain size distribution function which gives the number of dust grains per unit size per $H$ atom, $\mu(a)$ is the electric dipole moments as a function of grain size and $f_a(\omega)$ is the angular velocity distribution function which depends upon the grain radius and environmental condition. This function is calculated for a cold neutral medium, which corresponds to the case we adopt in our simulations.

In this work, we consider the standard model of the AME, constituting of simulated polarized maps with thermal dust polarization angles and nominal AME intensity. We exploit the model implemented in the {\tt Python Sky Model (PySM)} \footnote{https://github.com/bthorne93/PySM\_public} publicly available package which generates the full sky simulation in intensity and polarization \citep{pysm}. The AME model in {\tt PySM} makes use of {\tt SpDust2} code \citep{Ali_2008, Silsbee_2010} to calculate the nominal AME model and dust polarization angles ($\gamma_{353}$), are calculated from the Planck {\tt Commander} 2015 thermal dust Q and U maps at 353 GHz \citep{Adam:2015wua}. The assumption of a complete mixture of small and big grains leads to consider the same angles as thermal dust. The AME polarization can be written as:
\begin{equation}
\label{Ame}
Q_{ame} = f I_{\nu} \cos(2\gamma_{353}),\qquad U_{ame} = f I_{\nu} \sin(2\gamma_{353})\ , 
\end{equation}
where $f$ is the polarization fraction. In this work, we have considered a global 2\% for that, within the limits observed by WMAP, Planck, and QUIJOTE in Perseus.

\subsection{Simulations}
\label{sec:sim}
In this section, we describe the set up adopted to simulate the sky maps used to train and test our NN. As anticipated, we focus on the study of low frequency foregrounds, but we consider all the frequency channels covered by the future LiteBIRD satellite \citep{litebird}. Since we want to characterize low frequency foreground we also considered two additional low frequency channels. In particular we adopt the specification of the two lowest frequency bands of the QUIJOTE telescope with frequency of 11 and 13 GHz \citep{quijote}. Our choice is motivated as follows. The landscape of observations at low frequencies is evolving, and following the relevance of diffuse polarized foregrounds for $B$-modes, powerful low frequency observations are being proposed \citep[see e.g.][and references therein]{ground-LF}. In this work, we want to check how the present approach to foreground model recognition, based on NN, performs in the case of low frequency observations with detector technology which is available now, obtaining results which are, in this respect, conservative. The corresponding frequencies, together with sensitivities and angular resolutions for all the considered channels are summarized in Table \ref{table1}.\par

All of our sky simulations exploit the publicly available Hierarchical Equal Area Latitute Pixelization scheme for the sphere {(\tt HEALPix)}, and its Python package {(\tt healpy)}\footnote{https://github.com/healpy/healpy}, see \citep{healpix} for details.
The sky emissions included in our simulations are CMB, Galactic synchrotron, thermal dust and polarized AME. All the components are simulated using the {\tt PySM} and we refer to \citep{pysm} for details about how the different components are generated. \par
In particular, the CMB maps are generated as random Gaussian realizations of the Planck best fit $\Lambda$ Cold Dark Matter ($\Lambda$CDM) power spectrum \citep{planckI}. For dust, we have used the PySM template, rescaled as a modified blackbody, as in Equation (\ref{D1}), with constant spectral index and temperature $(\beta_d,\,T_d)=(1.54,\,20 K)$. For synchrotron, we have considered two different models. In the first one, the template is extrapolated in frequency with a simple power-law model. The spectral index is spatially varying, considering a Gaussian distribution with mean value $\beta_s$ = -3 and standard deviation equal to $0.2$. In the second case, a curvature is included in the synchrotron SED, with a constant value of $C=-0.052$, as indicated by Kogut \citep{kogut12} with 23 GHz as the pivot frequency; this setup is also compatible with the recent analysis by Krachmalnicoff et al. \citep{Krachmalnicoff2018}. Finally, as explained in the following Sections, we have also included, in some specific cases, the AME polarized signal, assumed to have a $2\%$ polarization fraction.\par

\begin{figure}
  \centering
  \includegraphics[width=.6\textwidth]{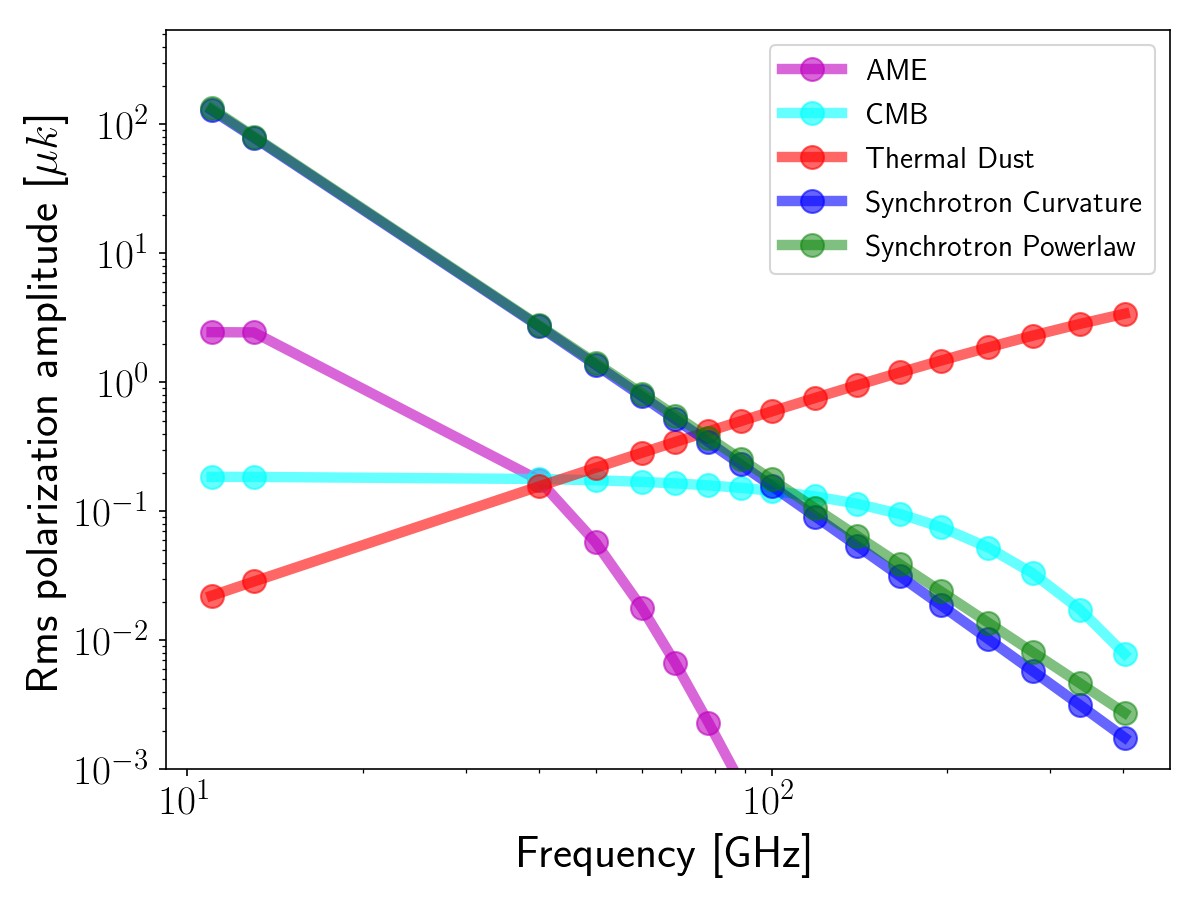}
  \caption{\label{fig:models} Polarized intensity standard deviation as a function of frequency for the different foreground components considered in this work. The plot refers to a sky fraction $f_{sky} = 78\%$ with $N_{side}=16$ HEALPix gridding (corresponding to pixels with side of about 4 degrees), and units are in brightness temperature.}
\end{figure}

In Table \ref{table2_models} we show a summary of the considered foreground models and the adopted parameterizations. As an illustration of the relative relevance of the various components, in Figure \ref{fig:models} we plot the standard deviation of their polarized intensity, in brightness temperature units and gridding the sky with 4 degree pixels, corresponding to the HEALPix gridding parameter  $N_{side}=16$, for all the sky emissions and frequencies considered in this work. To compute this plot we have applied Planck 2018 component separation common mask in polarization with $f_{sky} = 78\%$.

In our set up the noise is simulated uniformly in the sky, through Gaussian realizations with standard deviations given by the parameters listed in Table \ref{table1}.\par

\begin{table}
\begin{center}
\begin{tabular}{ |c|c|c|c| }
\hline
Experiment & Frequency [GHz] & Sensitivity [$\mu$K-arcmin] & FWHM [arcmin] \\
\hline
\multirow{2}{4em}{Quijote} & 11.0 & 840.0 & 55.2 \\
& 13.0 & 840.0 & 55.2 \\
\hline\hline
\multirow{15}{4em}{LiteBIRD} & 40.0 & 36.1 & 69.2 \\
& 50.0 & 19.6 & 56.9 \\
& 60.0 & 20.2 & 49.0 \\ 
& 68.0 & 11.3 & 40.8 \\
& 78.0 & 10.3 & 36.1 \\
& 89.0 & 8.4 & 32.3 \\ 
& 100.0 & 7.0 & 27.7 \\
& 119.0 & 5.8 & 23.7 \\
& 140.0 & 4.7 & 20.7 \\ 
& 166.0 & 7.0 & 24.2 \\ 
& 195.0 & 5.8 & 21.7 \\ 
& 235.0 & 8.0 & 19.6 \\
& 280.0 & 9.1 & 13.2 \\ 
& 337.0 & 11.4 & 11.2 \\ 
& 402.0 & 19.6 & 9.7 \\
\hline
\end{tabular}
\caption{Frequencies and instrumental specifications considered for preparing sky simulation in this work. The specifications are consistent with those of the LiteBIRD satellite and the QUIJOTE experiment
\citep{Errard2015}, \citep{Campeti:2019ylm}.}
\label{table1}
\end{center}
\end{table}

\begin{table}[]
    \centering
    \renewcommand{\arraystretch}{1.3}
    \begin{tabular}{|c|c|}
    \hline
    Foreground models  & Parameterization \\
    \hline
    Synchrotron power-law & $\mu({\beta_s}) = -3$, $\sigma({\beta_s}) = 0.2$ \\
    \hline
    Synchrotron curvature & $\mu({\beta_s}) = -3$, $\sigma({\beta_s}) = 0.2$, C = -0.052 \\
    \hline
    Thermal dust & $\beta_d = 1.54$, $T_d = 20K$ \\
    \hline
    AME & $f_p = 2$\% \\
    \hline     
    \end{tabular}
    \caption{Summary of the foreground models considered in this work. The parameterization is based on Equation \ref{S3} for synchrotron, Equation \ref{D1} for thermal dust and Equation \ref{Ame} for AME. $\mu({\beta_s})$ and $\sigma({\beta_s}$) are the synchrotron spectral index mean and standard deviation used to simulate the spatial variation of the parameters. $f_p$ represents the AME polarization fraction.}
    \label{table2_models}
\end{table}{}

\section{Neural Network architecture}
\label{sec:NNconcepts}

In this work, we have used NNs to recognize the actual parametrization of Galactic foregrounds in the sky. Generally speaking, NNs are algorithms that recognize underlying relationships in a set of data \citep{rashid2016}.
Given a function $f$, that maps an input $x$ into an output $y$, the goal of a NN is to find the best approximation $f^*$ of $f$. In order to do that, the NN recursively applies non-linear functions to linear combinations of the input elements. In this way, the function $f^*$ depends on several parameters $\theta$ (the coefficient of the linear combinations) which need to be optimized in order to get $f^*(\theta)\approx f$. This is done through a training set, i.e. a set of data for which the real output $y=f(x)$ is known: by computing the NN output $\tilde{y}$ for the elements of the training set, and by minimizing the distance between $y$ and $\tilde{y}$, the best values for the NN parameters $\theta$ are found. The optimization is done numerically, usually with a stochastic gradient descent method, and the function that encodes the distance between $y$ and $\tilde{y}$ is called \textit{loss function} \citep{nielsenneural}. \par 

There exist several NN architectures. In this work we make use of the so-called \textit{fully connected} ones. The basic structure of this kind of NN is a \textit{neuron}. Neurons are organized in layers; in each neuron a linear combination of all the elements of the previous layer is computed. These linear combinations are activated through a non-linear \textit{activation function}, and the outputs of this operation become the inputs of the following layer. In the {\it input layer}, neurons take the value of the elements of the input $x$, while in the output layer the neurons take the values of the elements of $\tilde{y}$. All the layers between the output and input ones are called {\it hidden layers}. For a general description of NN architectures can be found in \citep{goodfellow}.\par 

The set of $\theta $ values which constitutes the best approximation of $f$ is obtained through an iterative process, where the NN runs on the training set elements and the minimum of the loss function is found. The values of the $\theta$ parameters are updated at each \textit{epoch}. The number of epochs is one of the NN hyper-parameters and simply defines the number of iterations that are needed before the minimum of the loss function is reached. Given the very large number of parameters that a NN needs to optimize, over-fitting may occur; in this case, the NN approximates well the function $f$ on the training set but it is unable to generalize to another set of data. To avoid this, a typical approach is to introduce the so-called \textit{dropout}, i.e. a mechanism for which, in each epoch, some of the neurons of the NN are randomly switched off. This prevents the NN to rely on any specific parameter and allow it to mitigate overfitting. \par

We have built the NNs in the {\it Keras}\footnote{https://keras.io} environment, which is a {\it Python} library, with {\it Tensorflow}\footnote{https://www.tensorflow.org} backend. We have considered two NN architectures, which correspond to the problems we want to analyze, as described in the following.  

\subsection{NN for binary classification} 
\label{sec:BC_arch}

In our problem, the input of the NN are vectors of dimension $2 \times 17$. Each element of this vector represents the amplitude of the sky signal in a given pixel at the different considered frequencies (17 in total) for one of the polarization Stokes parameters. The two vectors of 17 elements each for Q and U are then stacked together to get the 34 elements long input vector.\par

As the purpose of this work is to solve a classification problem (assigning each pixel in the sky to a specific foreground model), the output of the NN is a vector where each element gives the probability that the input pixel belongs to any of the considered classes (models). The dimension of the output vector depends on how many possible sky models are considered, as explained in the following Sections.\par

\begin{figure}[]
  \centering
  \includegraphics[width=.5\textwidth]{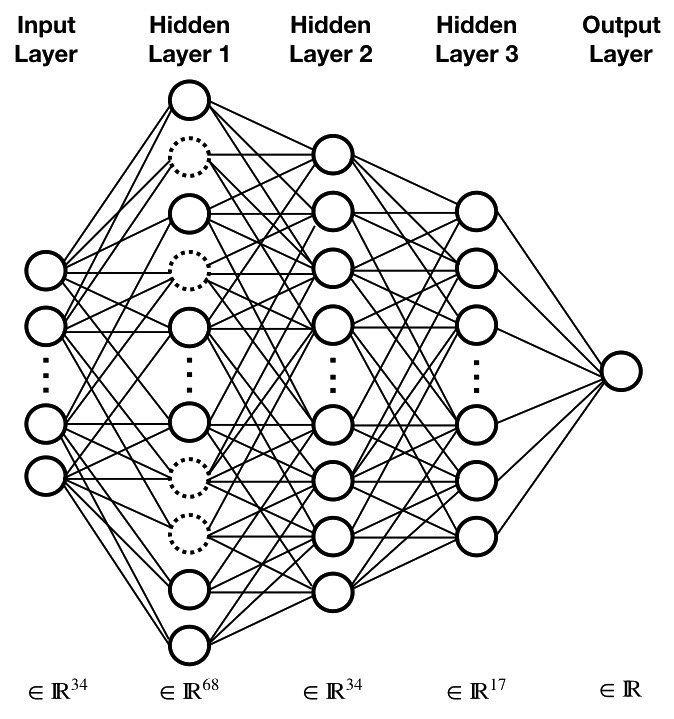}
  \caption{\label{fig:NN_arch}Schematic NN architecture used for binary classification: each circle represents a neuron, and the dashed circles indicate the application of dropout to a layer.}
\end{figure}

In the first considered case, we have trained the NN to perform a binary classification, meaning that its goal is to assign to each pixel in the sky one out of two possible foreground models.  As we specified above, the NN input layer has dimension of $34$, after that 3 hidden layers are present, including 68, 34, and 17 neurons each, with {\it tanh} as an activation function. In order to prevent overfitting, a dropout layer with a dropout rate = 0.5 is applied on the layer with the largest number neurons. Since we are in the case of binary classification, the output layer, activated with a \textit{sigmoid} function, has, in this case, dimensionality 1, corresponding to the probability of the input to belong to the first class. Figure \ref{fig:NN_arch} shows the schematic architecture of our binary classifier. The loss function is defined as a binary-crossentropy function: $L=−(zlog(p)+(1-z)log(1-p))$, where $p$ is the predicted probability for each input to belong to the specific class and $z$ is the binary indicator associated to the two classes (0 or 1). We have used \textit{Adadelta} optimizer with learning rate = 1.0 which is implemented in Keras.

\subsection{NN for multi-classification}
\label{sec:MC_arch}
As we explain in the following Sections, we have also considered a case where the NN has to distinguish among four different sky models. Due to the enhanced complexity with respect to the binary classification, we increase the number of layers and neurons accordingly. In this case, the NN has 5 hidden layers with 272, 136, 68, 34, and 17 neurons, with {\it tanh} activation function. As before, a dropout layer with a dropout rate = 0.5 is applied to the first hidden layer with 272 neurons. The output layer is a multi-classification, with {\it softmax} function as activation. A Sparse-Categorical-Cross-entropy is chosen as loss function, corresponding to $L = - \sum^{M}_{c=1} z_{o,c}log(p_{o,c})$, where $M$ is the number of categories for classification, $p$ is the  predicted probability for specific observation ($o$) of category $c$, and $z$ represents the correct class indicator for that observation ($o$). The same optimizer as the binary classification is considered.\par

\subsection{Hyper-parameters}
The values of the hyper-parameters describing the architecture of a NN apparatus is usually determined empirically. That is the case of the number of layers and the number of neurons per layer. A large number of these quantities ensure performance, at the expense of computational efficiency and speed. Usually, large values of hyperparameters are chosen and progressively reduced while keeping the performance stable, reaching minimum value which is then frozen in the NN apparatus. In our work, we have tried several NN configurations, and have selected, for both the cases of binary or multi-classification, the architecture which showed the best performance with the least number of parameters to be optimized during training. See \citep{goodfellow} and references therein for a general description of the hyper-parameter definition and derivation for NNs.


\section{Results}
\label{sec:res}

We now discuss the results of model recognition for low frequency foregrounds via NNs, both in binary and multi-model classification. The analysis is entirely based on simulated polarization maps, where the signal information is given via the $Q$ and $U$ Stokes parameters. We study noisy and noiseless simulated maps at LiteBIRD and QUIJOTE frequencies, as anticipated. In the following Sections we describe the results for the different test cases we consider.\par


\subsection{Foreground model recognition via binary classifications}
\label{sec:BC}

We first use the binary classifier described in Section~\ref{sec:BC_arch} to distinguish between two different foreground models. In particular, in the first case we train the NN in order to understand whether low frequency data are fitted better by a synchrotron model which does or does not include curvature of the spectral index (see Equation \ref{S3}). Next, we focus on the case in which the synchrotron emission is described by a pure power-law and the NN is trained to recognize the presence of polarized AME.

\begin{figure}[]
\centering 
\includegraphics[width=.5\textwidth]{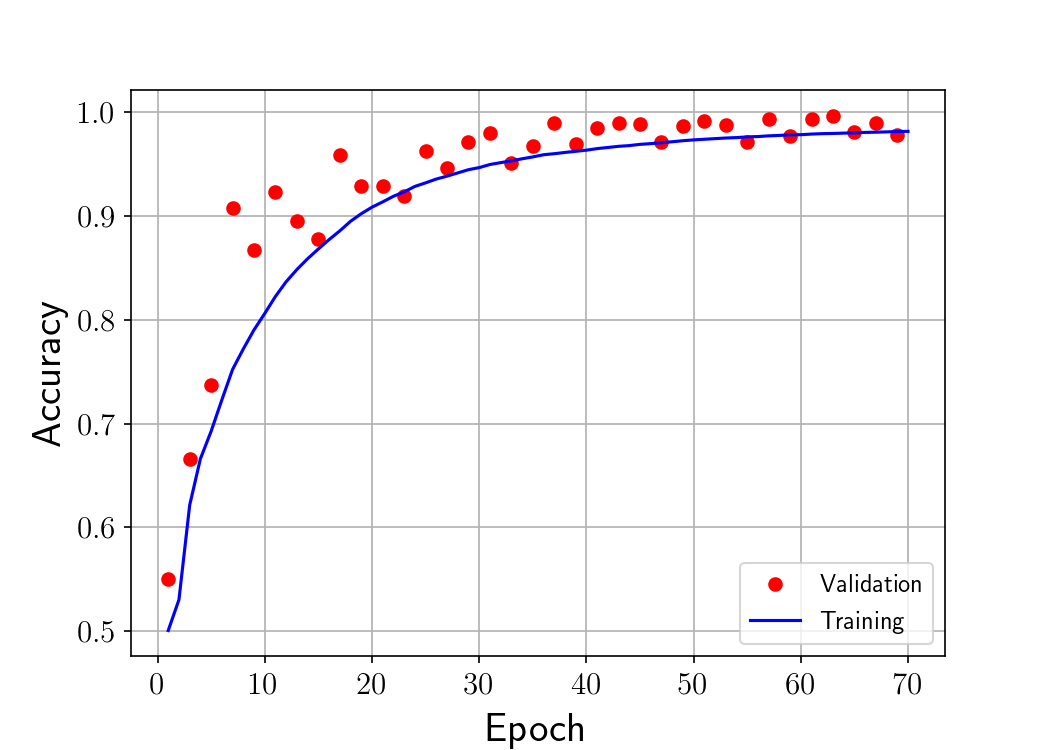}
\caption{\label{fig:S1S3_acc} NN accuracy as a function of the training epochs for binary classification between synchrotron with and without curvature, in the noiseless case.}
\end{figure}

\subsubsection{Synchrotron curvature}
\label{sec:BC_synch}

We have trained the NN with four sets of simulated multi-frequency maps. Each set of maps consists in 34 skies, i.e. 17 frequencies for Stokes $Q$ and $U$ emissions. In each set we have included the emission coming from the CMB, polarized thermal dust and synchrotron simulated as described in Section \ref{sec:sim}. In two sets of maps, the synchrotron emission is scaled in frequency with a pure power-law, while in the remaining two a curvature is added to the spectral index. We have considered a different random realization of the CMB emission for each set of maps, as well as a different realization of the synchrotron spectral index spatial variation, which is taken from a Gaussian distribution with mean -3 and standard deviation 0.2. 
The synchrotron curvature in the two sets of maps is constant, with a value of $C=-0.052$, and 23 GHz as the pivot frequency. All the maps have been simulated at $N_{side} = 1024$, meaning that in total we have about $5\times10^7$ vectors, each of which consists of 34 elements, which are used for training. Among these, we have randomly selected 20\%, which are not used for optimizing the NN weights, but as a validation set, as it is typically done for validating the performance of a NN. The size of the training set has been chosen in order to find the optimal balance between NN performaces and computational costs. Since we have considered all the pixels in the sky maps for training the NN, and given the high level of non-stationarity of the Galactic signals, the vectors in the training set cover a very large dynamic range, of about four orders of magnitude. As it is done in preparing the data for NN training, we normalized each input vectors in the range between -1 and 1 as follows: the minimum and maximum value for each input vector are computed; the minimum value is subtracted to the vector elements, and the result is divided by the difference between maximum and minimum. 
In order to further generalize the training set and make it substantially different from the test one, we have shifted the amplitude of each Galactic component. In particular, we have applied a multiplication to both the templates of synchrotron and thermal dust (at 23 and 353 GHz respectively): in each template, each pixel in $Q$ and $U$ is multiplied by a random value drawn from a Gaussian distribution with standard deviation equal to 30\% of the amplitude of the pixel itself. The multi-frequency maps are then obtained by applying the correct frequency scaling to these modified templates.\par

In Figure \ref{fig:S1S3_acc} we show the training history with the accuracy reached by the NN as a function of epochs. Since we are working on a classification problem, in this case the accuracy represents the percentage of elements in the training (or validation) set which are classified correctly. We recall that the NN outputs the probability for each input pixel to belong to each considered class and that each pixel is assigned to the class that has the highest probability.\par

\begin{figure}[]
\centering
\includegraphics[width=.6\linewidth]{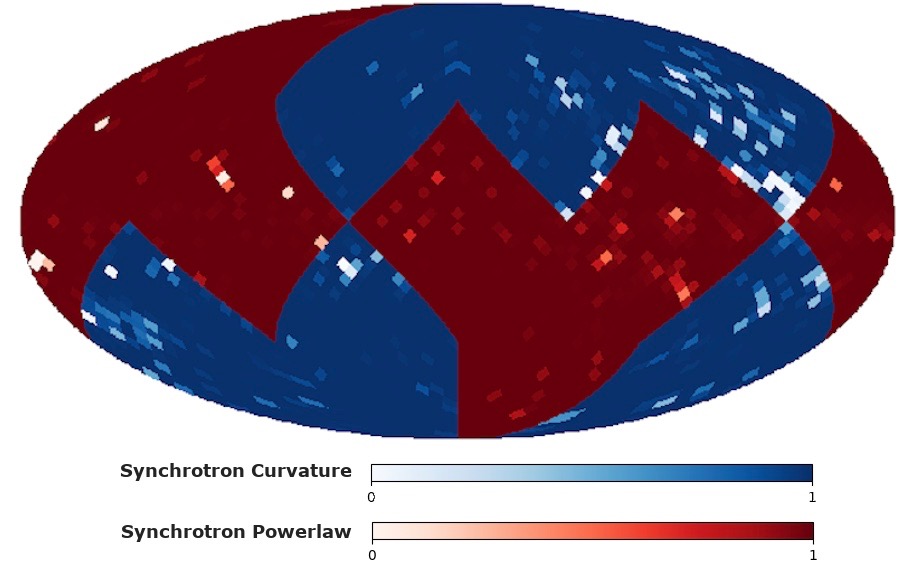}
\hspace{0.1cm}
\raisebox{0.99\height}{\includegraphics[width=.3\linewidth]{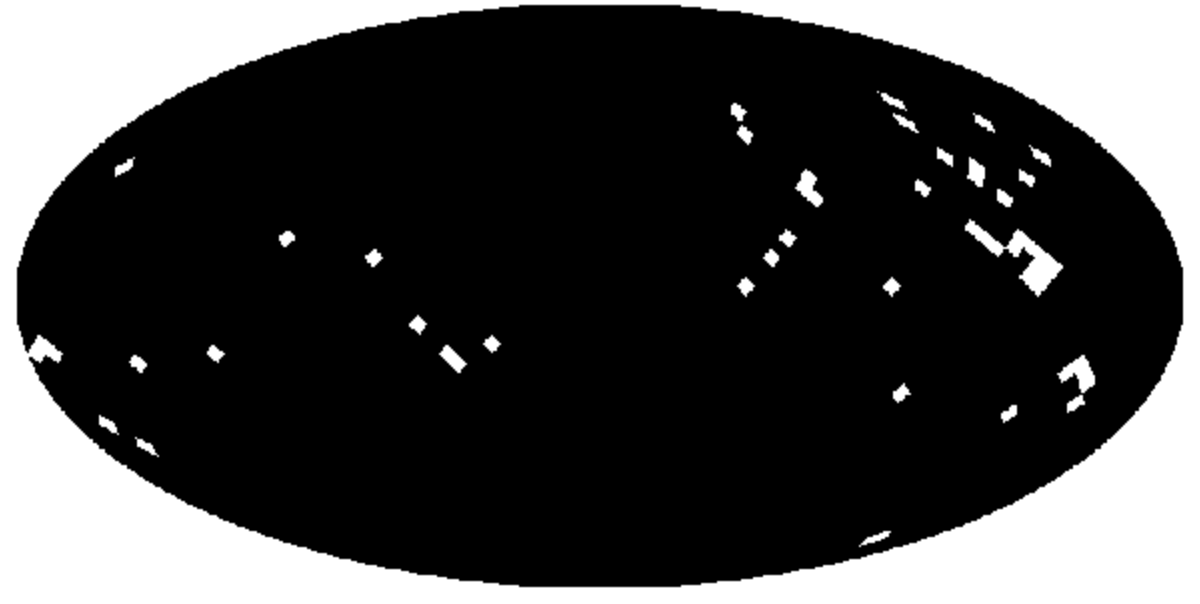}}
\caption{\label{fig:S1S3_predict} Left panel: NN prediction on a test set map for binary classification of the Galactic synchrotron with (blue regions) and without curvature (red regions) in the ideal case of noiseless maps. The color bar shows, for each pixel, the probability to belong to the correct class, as assigned by the NN. Lighter pixels are those where the incorrect model has been assigned. Right panel: for sake of clarity, correct (black) and incorrect (white) pixels are also shown with a binary color scale.}
\end{figure}

\begin{figure}[]
\centering 
\includegraphics[width=.55\textwidth]{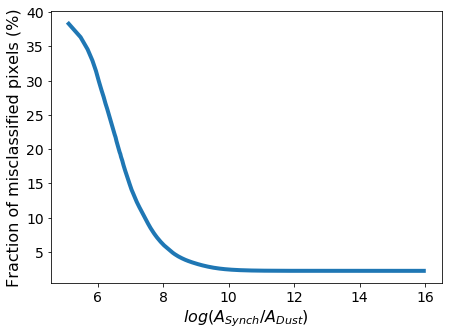}
\caption{\label{fig:S1S3_prob} Faction of misclassified pixels as a function of the relative amplitude of synchrotron over dust at 11 GHz for the case of binary classification of the Galactic synchrotron with or without curvature.}
\end{figure}

Once the NN is trained, we can apply it to the test set. In particular, we have built test maps, by making use of the PySM library, that include CMB, synchrotron, and thermal dust. 
Maps of the test set have been generated at $N_{side}=16$ and without the modulation of the foreground templates in order to make our test set considerably different from the training one. In some regions, the synchrotron emission has been scaled in frequency with a simple power-law, in others, we have 
modified the SED by including a running parameter of the spectral index. An example of a test set map is reported in Figure \ref{fig:S1S3_predict}: in the pixels belonging to the red regions the synchrotron SED is a pure power-law, while in the blue region a curvature is added. The color scales in Figure \ref{fig:S1S3_predict} report the output of the NN, i.e. the probability that each pixel belongs to the correct class. In particular, pixels shown with darker colors are those where the NN assigned the correct class, while pixels with lighter colors are those where the NN has missed the right foreground model. For sake of clarity, in the right panel of Figure \ref{fig:S1S3_predict}, we show, in white, the pixels where the NN has made an incorrect prediction. The achieved accuracy (i.e. the percentage of correctly classified pixels) is about 98\%. We have tried different combinations of patterns for synchrotron power-law and curvature in the sky, assessing that the accuracy reached by the NN is stable and does not depend on the considered sky configuration. 
\\
We have also investigated the physical properties of those pixels where the NN assigned the wrong model. In particular, we have found that when the relative amplitude of the synchrotron emission over dust is small, 
the NN has the tendency to misclassify the model. This happens for example in the region near Galactic coordinate (230$^{\circ}$, +40$^{\circ}$) where the synchrotron amplitude is known to be extremely weak, or on the Galactic plane where dust emission is very bright.
We have quantified this effect in Figure \ref{fig:S1S3_prob}, where we show the fraction of misclassified pixels as a function of the relative amplitude of synchrotron over dust emission. In particular, we have considered a map at $N_{side}$ = 256 (corresponding to about $7.8\times10^5$ pixels) where we have scaled the synchrotron emission with a pure power-law on the whole sky. For each pixel, we have computed the synchrotron over dust amplitude at the frequency of 11 GHz and for the total polarized intensity. We have applied a binning on this ratio such that in each bin we have the same number of pixels (about 1600). A threshold corresponds to each bin, and we have counted the ratio of misclassified pixels over the total number of pixels with $\log(A_{synch}/A_{dust})$ below the threshold. The results in Figure \ref{fig:S1S3_prob} show that when the synchrotron over dust amplitude is small, the fraction of misclassified pixels increases, up to about 38\%, while for the pixels where synchrotron emission is high compare to dust, the faction of misclassified pixels decreases dramatically.

\subsubsection{Synchrotron and AME}
\label{sec:BC_ame}

We have used the same NN architecture developed for binary classification with the goal of identifying those pixels where AME polarized radiation is present in the sky. The two models considered in this case correspond therefore to Galactic synchrotron with a pure power-law SED, or synchrotron plus polarized AME component with the specifications described in Section \ref{sec:sim}. 

\begin{figure}[]
\centering 
\includegraphics[width=.6\textwidth]{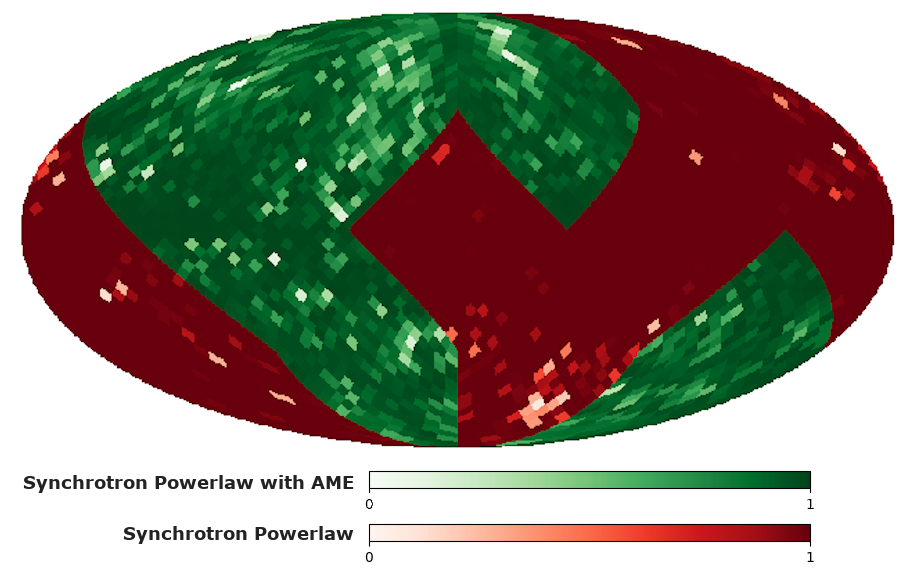}
\hspace{0.1cm}
\raisebox{0.99\height}{\includegraphics[width=.3\textwidth]{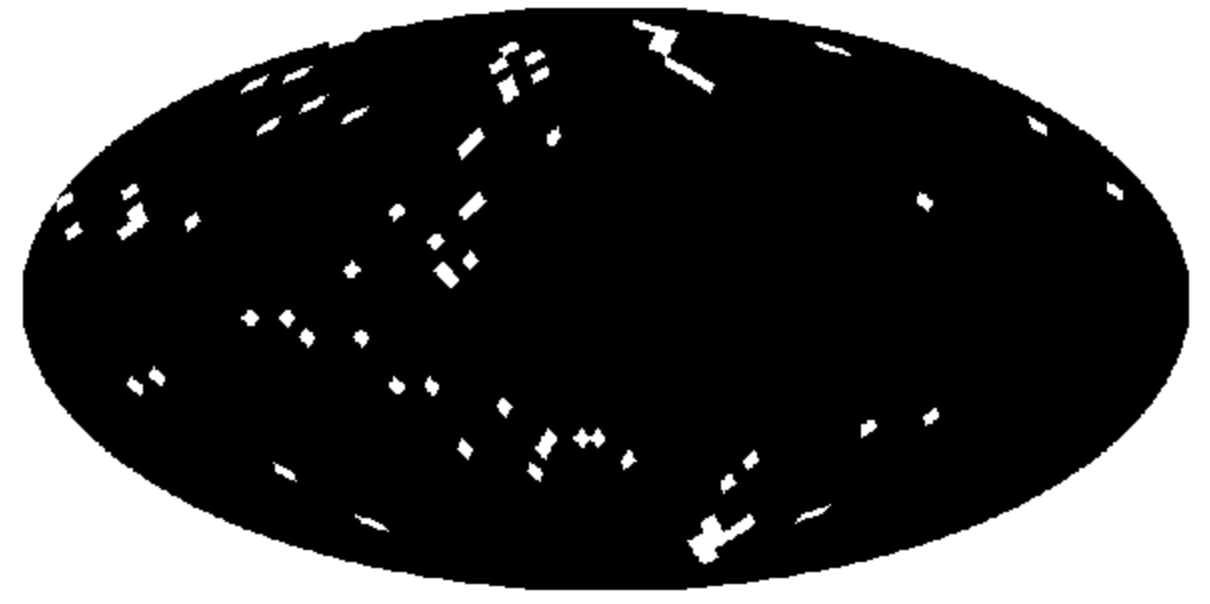}}
\caption{\label{fig:AMES1_predict} Left panel: NN prediction on a test set map for binary classification for a sky emission that at low frequency includes synchrotron with (green regions) or without (red regions) AME. The color bar shows the NN probability assigned to the correct model. Right panel: white pixels indicate regions where the NN has assigned the incorrect model. These results are for the noiseless case.}
\end{figure}

\begin{figure}[]
\centering 
\includegraphics[width=.55\textwidth]{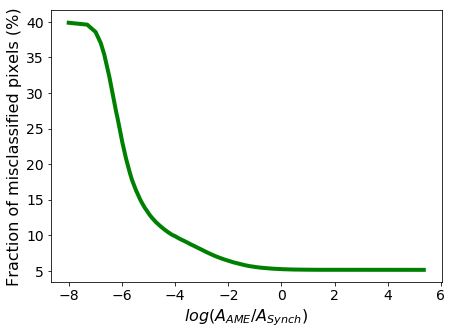}
\caption{\label{fig:AMES1_prob} Fraction of misclassified pixels as a function of the relative amplitude of AME over synchrotron at 40 GHz, for the case of binary classification of the Galactic synchrotron with or without AME.}
\end{figure}

For what concerns the training, we have followed a procedure analogue to the one presented in the previous Section. The training consists of four sets of maps; in two of them we have simulated the sky emission by considering the presence of CMB, synchrotron and thermal dust radiation, while in the remaining two we have also included polarized AME. As before, the total number of vectors in the training set is about $5\times10^7$ and the templates of foreground emissions (dust, synchrotron and AME) have been modified by applying the multiplication factor as described in the previous Section. Results are presented in Figure \ref{fig:AMES1_predict}, where AME is present are shown in green. In the ideal noiseless case, the NN is able to correctly classify the foreground model in about 97\% of the cases. We highlight that pixels where the NN fails in classifying correctly the foreground models are those where the AME emission is faint with respect to the synchrotron one.
In Figure \ref{fig:AMES1_prob} we report the fraction of misclassified pixels as a function of the relative amplitude of AME over synchrotron at 40 GHz (the frequency closest to the AME peak), similarly to what we have done for Figure \ref{fig:S1S3_prob}. The results show that, as expected, the smaller AME amplitude is compared to synchrotron, the higher is the fraction of misclassified pixels, up to about 40\%. \par

\subsection{Multi-model classification}
\label{sec:MC}

We now extend the study performed so far and consider a more complex case in which the NN is trained to classify four different foreground models in the simulated sky. In this case we have used the NN architecture described in Section~\ref{sec:MC_arch}. As before, we have built our simulated maps by including CMB and thermal dust, while the low frequency foregrounds include synchrotron with or without a curved SED and possibly AME.\par

\begin{figure}[]
\centering 
\includegraphics[width=.5\textwidth]{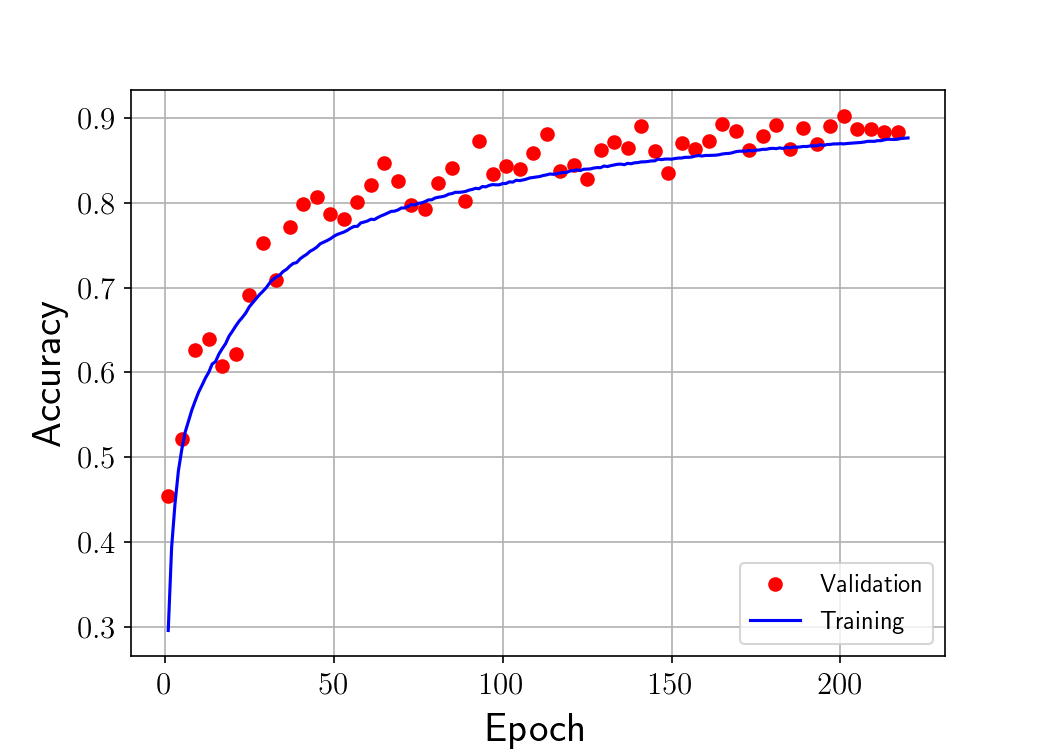}
\caption{\label{fig:AMES1S3_acc} NN accuracy as a function of the training epochs for multi-class classification between synchrotron with or without curvature and with or without AME, in the noiseless case.}
\end{figure}

The training set has been generated from four sets of maps as before, for a total of about $5\times10^7$ vectors used for optimizing the NN weights. The training history is shown in Figure \ref{fig:AMES1S3_acc}: the NN reaches about 87\% of accuracy on the training set after 220 epochs. It is worth noticing that as a result of the enhanced complexity in the simulations, the NN training takes more time to optimize weights, reaching convergence in about 220 epochs.\par

\begin{figure}[]
\centering
\includegraphics[width=.6\textwidth]{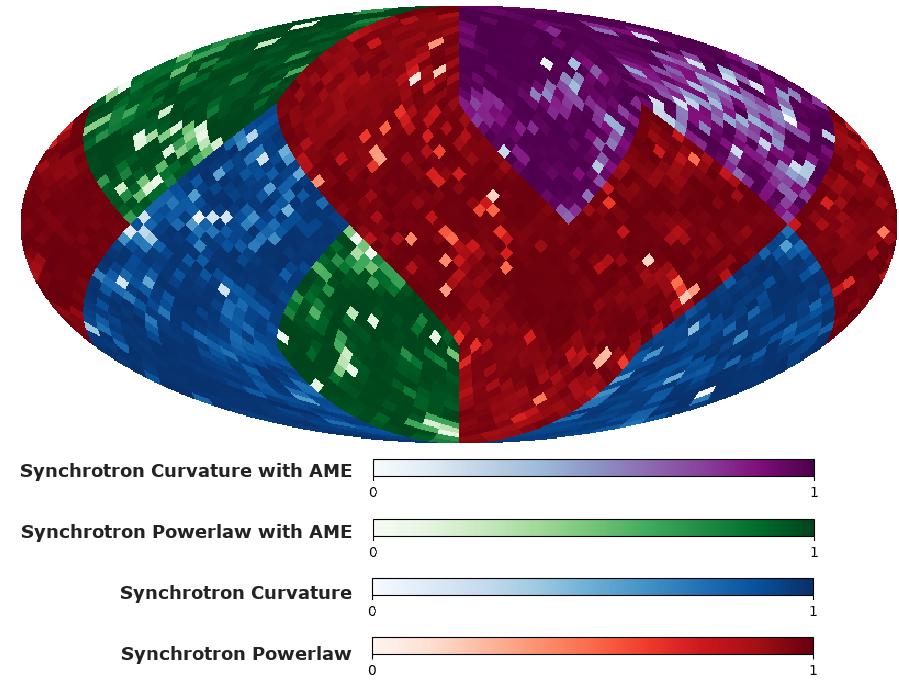}
\hspace{0.1cm}
\raisebox{1.48\height}{\includegraphics[width=.3\textwidth]{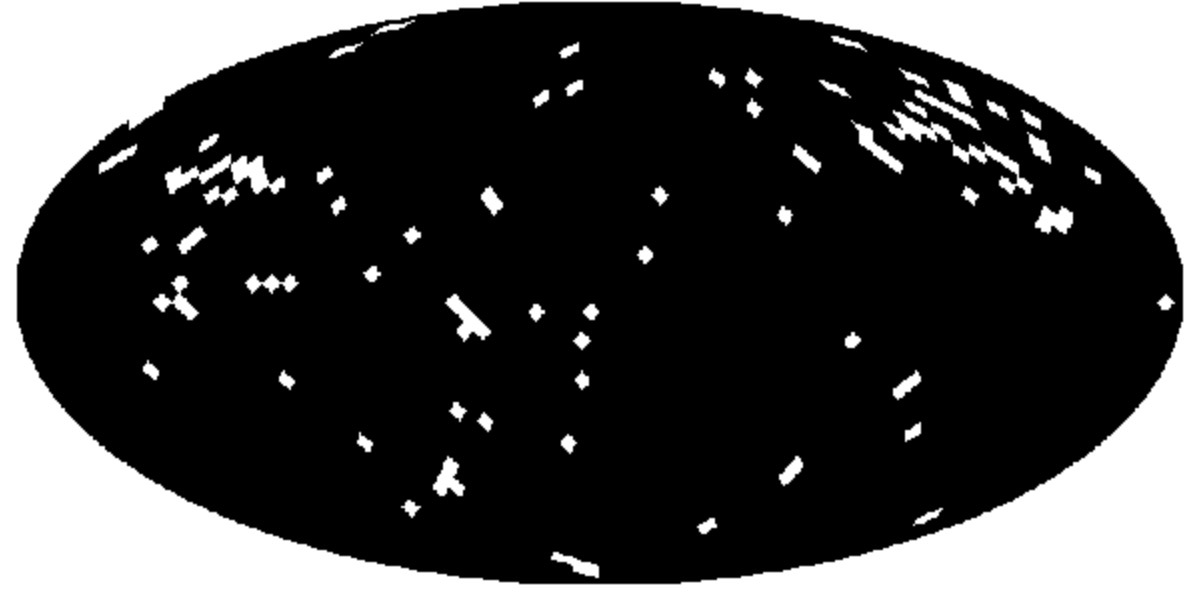}}
\caption{\label{fig:AMES1S3_predict} Left panel: NN prediction on a test set map for the multi-class classification for pure power-law synchrotron with (green regions) and without AME (red regions), or for synchrotron with curvature with (purple) and without (blue) AME. The color bars show the NN probability assigned to the correct model. Right panel: white pixels are those where an incorrect model is indicated by the NN. These results are for the noiseless case.}
\end{figure}

Results on a test map are shown in Figure \ref{fig:AMES1S3_predict}. In this case the sky is divided into four different regions, corresponding to the four models that the NN has to classify: synchrotron with a pure power-law SED (red), synchrotron with running of the spectral index (blue) and presence of polarized AME (green when AME is added to the synchrotron power-law model and purple when it is added to synchrotron with curvature). As before, color bars report the probability obtained by the NN that a given pixel belongs to the correct class, with lighter colors showing pixels where the NN has been assigned with the incorrect foreground model. The reached accuracy on the test set is at the level of about 93\% and as before it does not depend on the specific pattern of models in the sky.\par

In Table~\ref{table_acc_without_noise} we report a summary of the performance of the NN in the different considered configurations. We notice that in some cases the accuracy reached on the test set is higher than the one on the training set, as it could happen as a consequence of having exploited dropout during training.  

\begin{table}[]
    \centering
    \begin{tabular}{|p{7.2cm}|p{2.2cm}|p{2.2cm}|}
    \hline
    Sky models  & Accuracy on training set & Accuracy on test set \\
    \hline
    Pure power-law \& Curvature & 99\% & 98\% \\
    AME \& Pure power-law & 93\% & 97\% \\
    AME \& Pure power-law \& Curvature & 87\% & 93\% \\
    \hline     
    \end{tabular}
    \caption{Accuracy on training and test sets of the NN for different sky models in the basic configuration without noise.}
    \label{table_acc_without_noise}
\end{table}{}

\subsection{Classification in presence of noise}
\label{sec:MC}
We have tested the performances of our NNs when instrumental noise is present on maps. In particular, we have considered the specification of the LiteBIRD and QUIJOTE experiments, with the sensitivities reported in Table \ref{table1} and uniform white noise distribution across the sky.\par

Our first approach has been to change only the test sets, by adding noise on the test maps, but keeping the weights of the NNs unchanged, therefore with the values optimized with the noiseless training. The first column of Table \ref{table3} reports the accuracy reached on the test sets for the three classification schemes we considered: binary classification for synchrotron models, presence of AME, and multi-classification. For the binary classification, we reached acceptable accuracy; While the accuracy drops significantly, reaching about 68\% in the more complex multi-classification case.\par

In order to get better results, we have trained the NN with noise in the training set. We have considered two different approaches. In the first one, we have added one noise realization on the multi-frequency maps used previously as the training set. We have then taken the NN trained previously on noiseless data, and performed a second phase of training with the noisy training set. In this way, the NN shows a remarkable improvement in accuracy, being able to reach $\sim 90\%$ on the test set for the multi-classification. In the second  approach, we have built new training sets, consisting in 100 maps for each model at low resolution ($N_{side}$ = 16), resulting in 400 sets of maps included in the training set, corresponding to more than 1 million pixels. Similarly to the previous case, the accuracy is pretty high, at the level of about 93\%, proving that, during training, the NN is able to learn the noise properties and take those into account during the model classification.\par

\begin{figure}[]
\centering
\includegraphics[width=.6\textwidth]{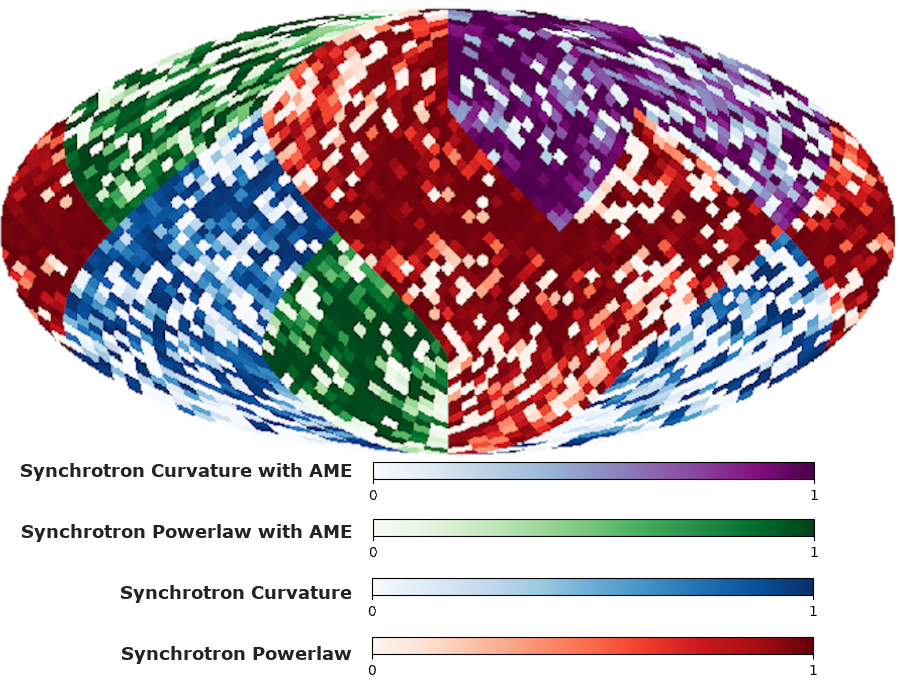}
\hspace{0.1cm}
\raisebox{1.4\height}{\includegraphics[width=.3\textwidth]{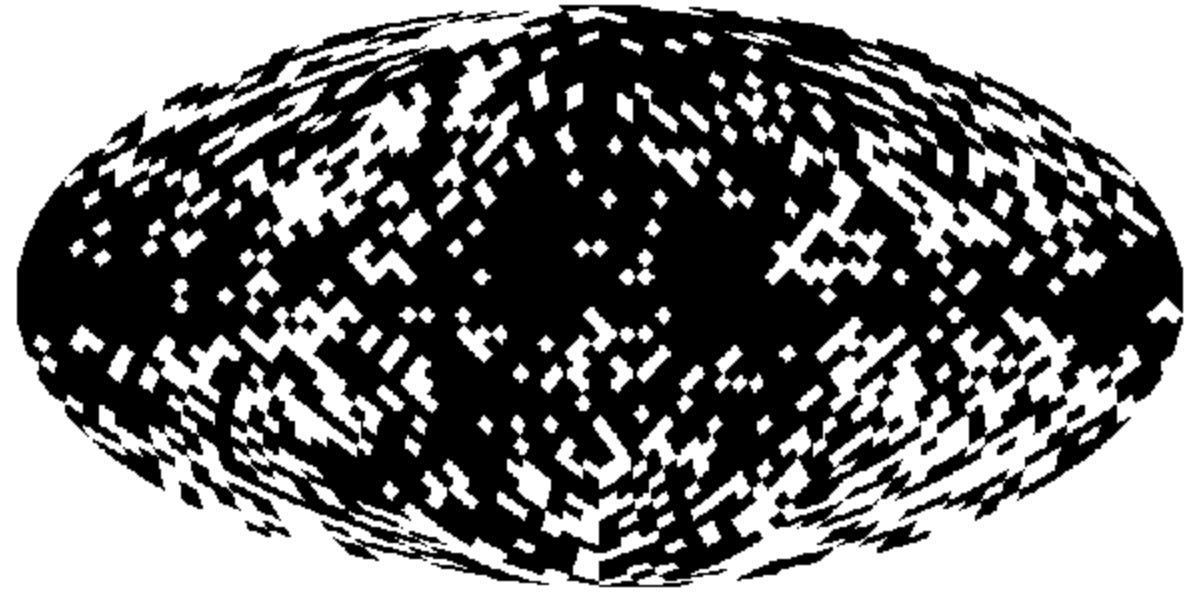}}
\\
\includegraphics[width=0.6\textwidth]{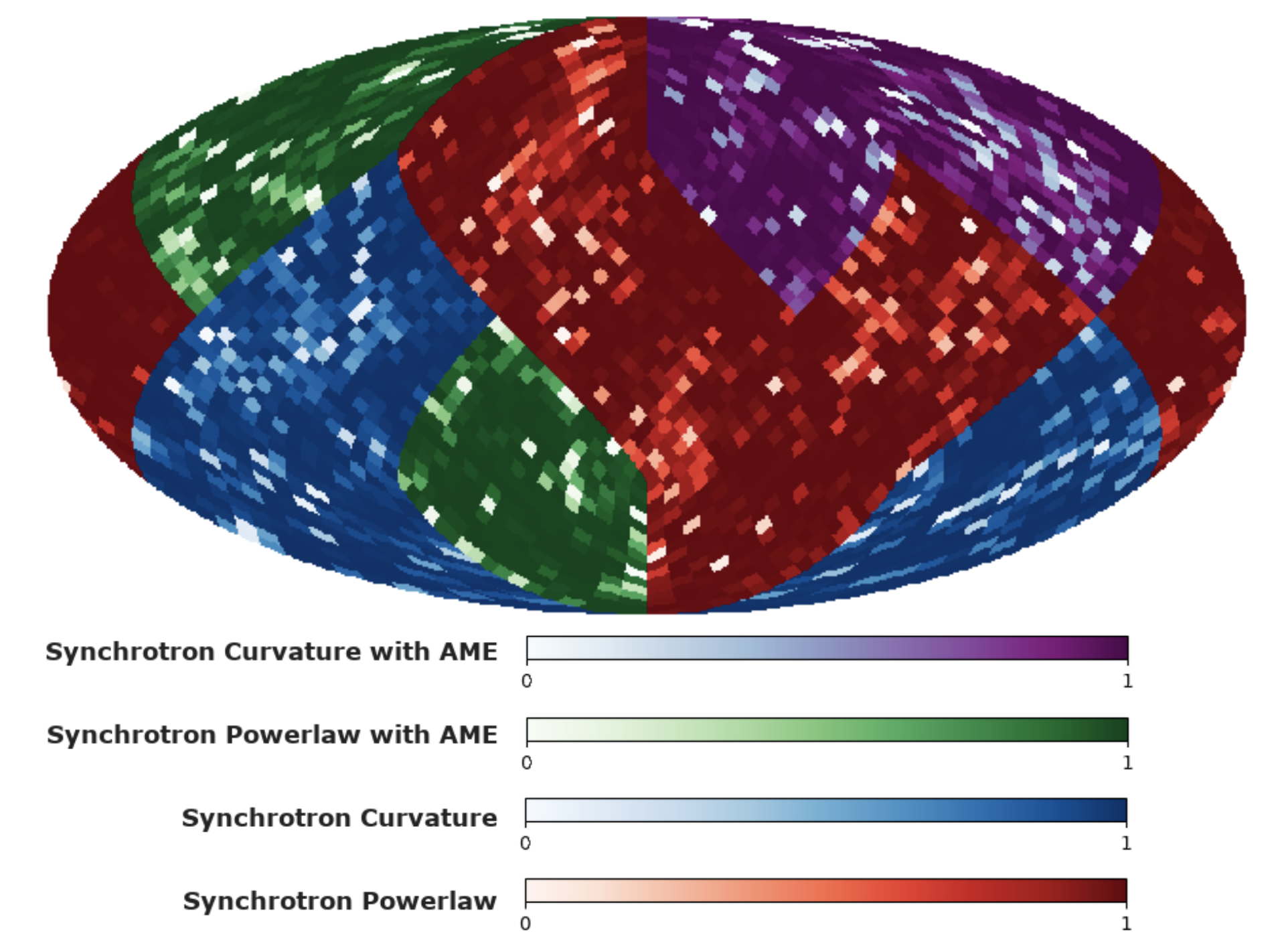}
\hspace{0.1cm}
\raisebox{1.35\height}{\includegraphics[width=0.3\textwidth]{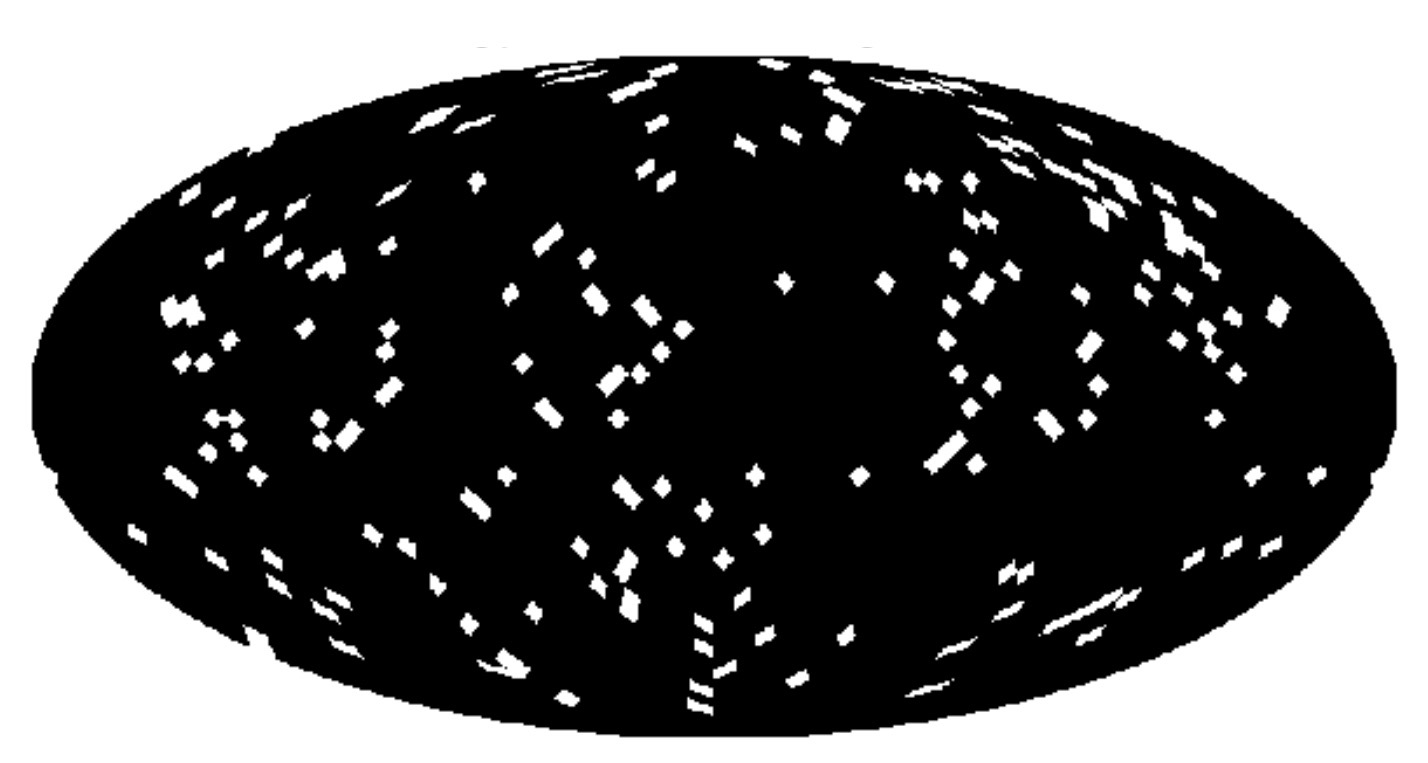}}
\caption{Effect of including the presence of noise in the training set. The color scales for the considered models are the same as Figure \ref{fig:AMES1S3_predict}. The upper panels indicate the NN accuracy when noise is added only to the test set, lower panels show the NN accuracy on the same noisy test set after re-training the NN with 100 noise realizations at $N_{side}$ = 16; As before, white pixels in the right panels are those where the incorrect model is indicated by the NN.}
\label{fig:AMES1S3_noisy}
\end{figure}

In Figure~\ref{fig:AMES1S3_noisy} we show the results on the noisy test map for the multi-model classification, for the case in which the training has been done with noiseless simulations (upper panels) and the one where the training set was obtained from low resolution maps (lower panels). A summary of all the results is reported in Table \ref{table3}.

%

\begin{table}[]
    \centering
    \begin{tabular}{|p{3.3cm}|p{3.5cm}|p{3.5cm}|p{3.5cm}|}
    \hline
    \begin{center}Sky models\end{center} & Test set accuracy for noiseless data training & Re-training accuracy with $N_{side}=1024$, 1 noise realization & Re-training accuracy with $N_{side}=16$, 100 noise realizations\\
    \hline
    Pure power-law \& Curvature & \begin{center}82\%\end{center} & \begin{center}95\%\end{center} & \begin{center}97\%\end{center} \\
    \hline
    AME \& Pure power-law & \begin{center}78\%\end{center} & \begin{center}92\%\end{center} & \begin{center}94\%\end{center} \\
    \hline
    AME \& Pure power-law \& Curvature & \begin{center}68\%\end{center} & \begin{center}90\%\end{center} & \begin{center}93\%\end{center} \\
    \hline     
    \end{tabular}
    \caption{Accuracy of the NN for the binary and multi-model classification in the presence of noise with different approaches for training.}
    \label{table3}
\end{table}{}


\section{Comparison with the $\chi^{2}$ information}
\label{sec:comparison}

In this Section, we compare quantitatively the information retrieved via our NN apparatus with the ordinary goodness of fit represented by a $\chi^{2}$ test following a parametric component separation analysis. Here we exploit the approach developed by Stompor et al. \citep{composep} which is currently used for quantifying the science outcome of future $B$-mode probes \citep{Campeti:2019ylm}. We refer to these papers for further details on this approach, limiting ourselves to the definition of quantities of relevance for the present work. The data model is usually written as 
\begin{equation}
\label{dp}
    d_p(\nu) = \sum_{c} {a_p^c(\nu)}{s_p}^{c} + n_p(\nu) \equiv \mathbf{A}_{p}s_p + n_p\,,
\end{equation}
where $d_p$ contains measured signal at each frequency $\nu$ and sky direction $p$, summed over all components whose amplitude is written as $s_{p}^{c}$; $\mathbf{A}_{p}$ is the mixing matrix which contains the parametric SED model to fit, depending in principle on the sky direction, and $n_p$ represents the noise. The component separation process consists in obtaining an estimate $\tilde{s}_{p}=\mathbf{W}_{p}d_{p}$ of the components, by means of a kernel operator $\mathbf{W}_{p}$, given by
\begin{equation}
\label{chi2}
    \mathbf{W}_{p} \equiv (\mathbf{A}_{p}^T\mathbf{N}_{p}^{-1}\mathbf{A}_{p})^{-1} \mathbf{A}_{p}^T \mathbf{N}_{p}^{-1}, \qquad \mathbf{N}_{p} \equiv {n_p}^T n_p\,,
\end{equation}
where $\mathbf{N}_{p}$ represents the noise correlation matrix; the kernel operator is the result of the maximization of the likelihood 
\begin{equation}
\label{likelihood}
    -2\log\mathcal{L} = -\sum_p ( d_{p}-\mathbf{A}_{p}s_{p} )^T\mathbf{N}_{p}^{-1} ( d_{p}-\mathbf{A}_{p}s_{p} )\,,
\end{equation}
which is valid in the case in which the noise is block diagonal, i.e. correlations are allowed between Stokes parameters in a given pixel only. 
The $\chi^{2}$ is defined as 
\begin{equation}
\label{chi2}
    \chi^2(p) = \sum_{\nu=1}^{N_{band}}\left(\frac{d_{\nu} - {s}_{\nu}(p)}{\sigma_{\nu}(p)}\right)^2,
\end{equation}
Where $\sigma_{\nu}(p)$ represents the uncertainty due to the presence of noise. The corresponding approach to component separation has been implemented into the publicly available code called ForeGround Buster ({\tt FGBuster})\footnote{ https://github.com/fgbuster/fgbuster}, which we adopt in the rest of the work for calculating the $\chi^{2}$ after component separation, using the same input maps used so far for the NN. We restrict this analysis to the classification in the simplest cases of pure power-law or curved SED for synchrotron, i.e. the first case analyzed in the previous Section, in the binary classification mode. We run {\tt FGBuster} on the skies used to test the NN in the presence of noise, and calculate the $\chi^{2}$ accordingly. For all the pixels we fit two different models: in one case, the parameters to fit with {\tt FGBuster} are synchrotron, dust amplitudes and synchrotron spectral index, while in the other case, in addition to those, we also fit for synchrotron curvature. Since the parameterization of two synchrotron models is different, in order to have a fair comparison between the two $\chi^{2}$ tests, we have computed the reduced $\chi^{2}$ taking into account the degrees of freedom.\par

\begin{figure}[]
\centering
\includegraphics[width=.6\textwidth]{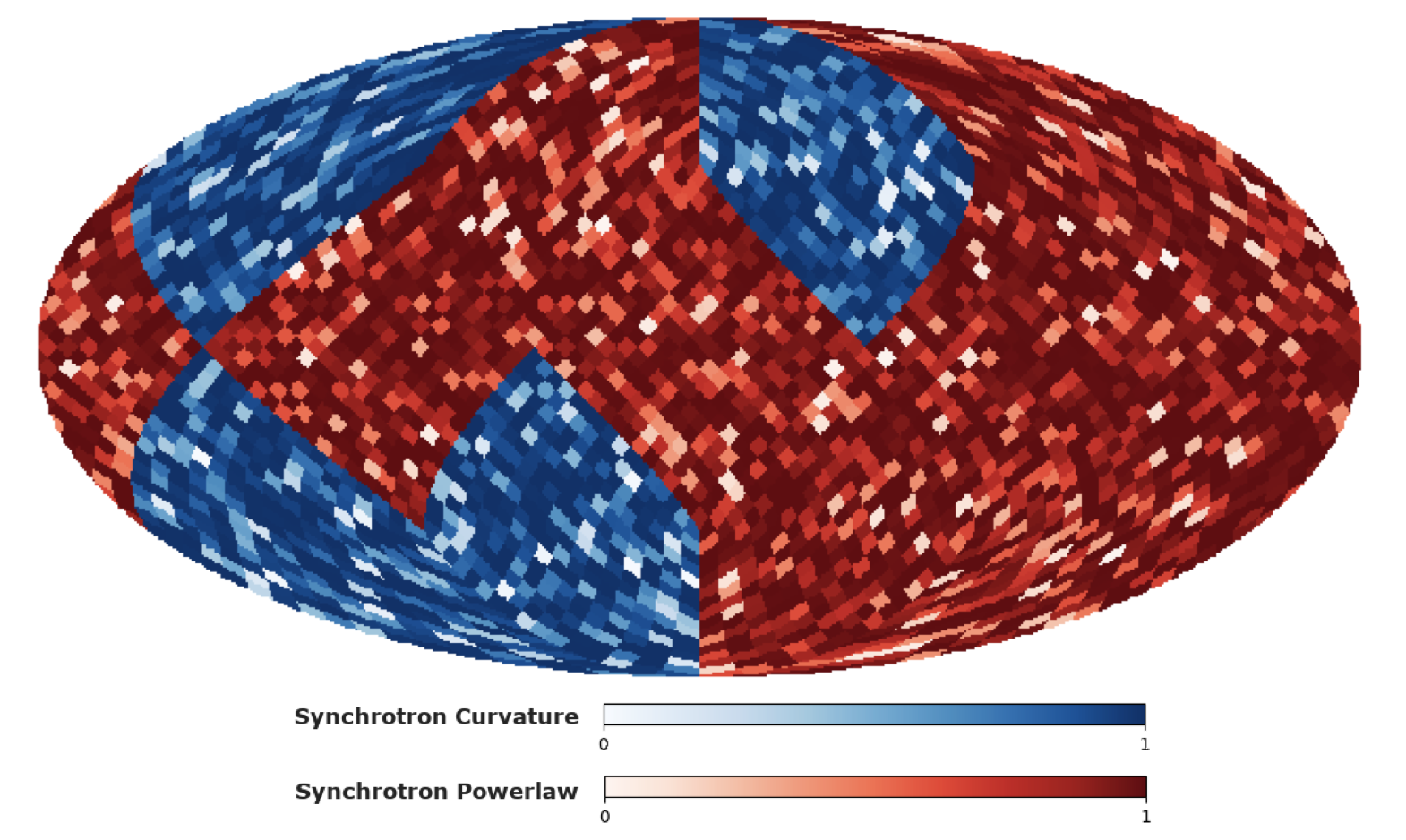}
\hspace{0.1cm}
\raisebox{0.88\height}{\includegraphics[width=.3\textwidth]{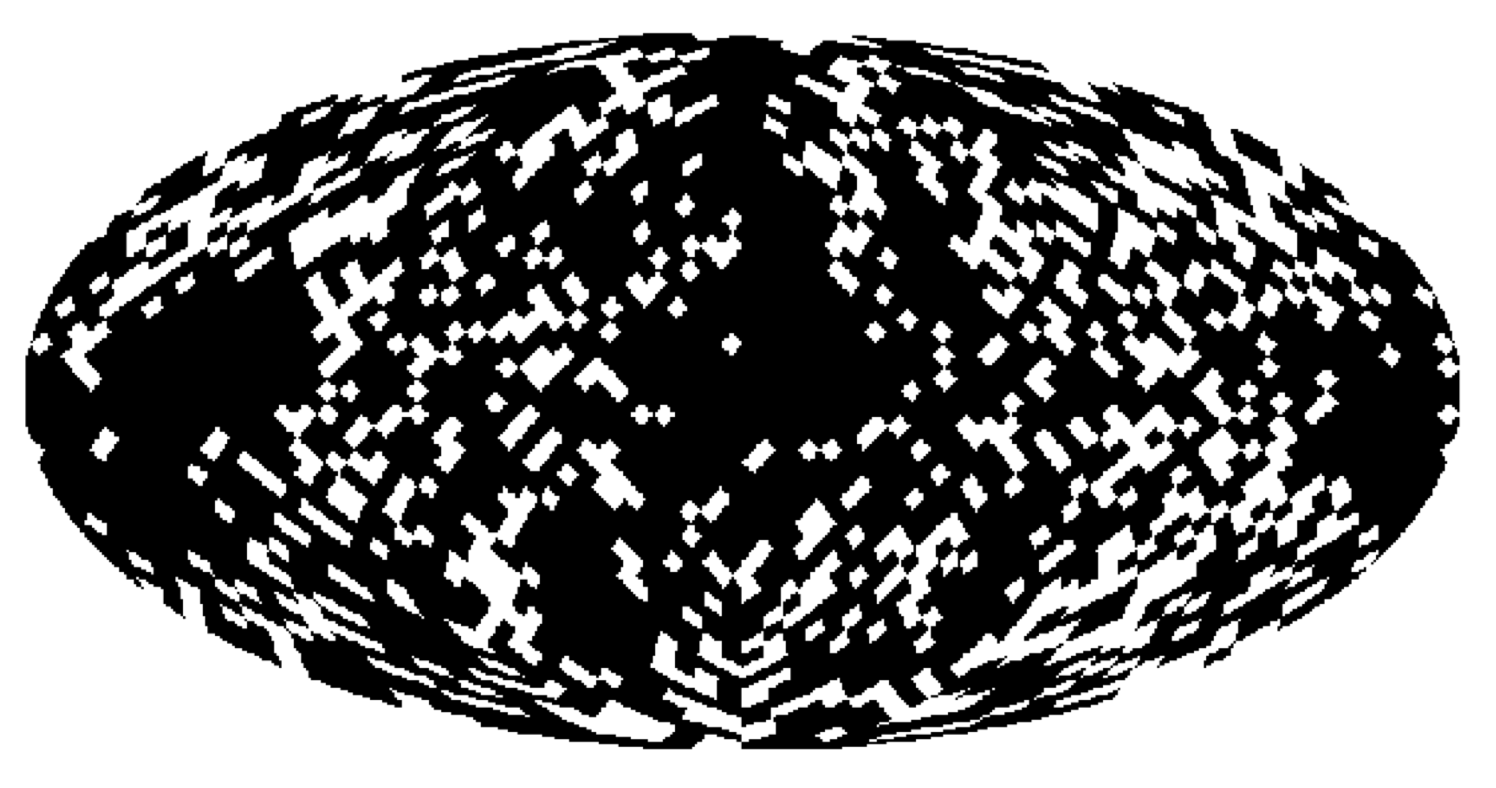}}
\\
\includegraphics[width=0.6\textwidth]{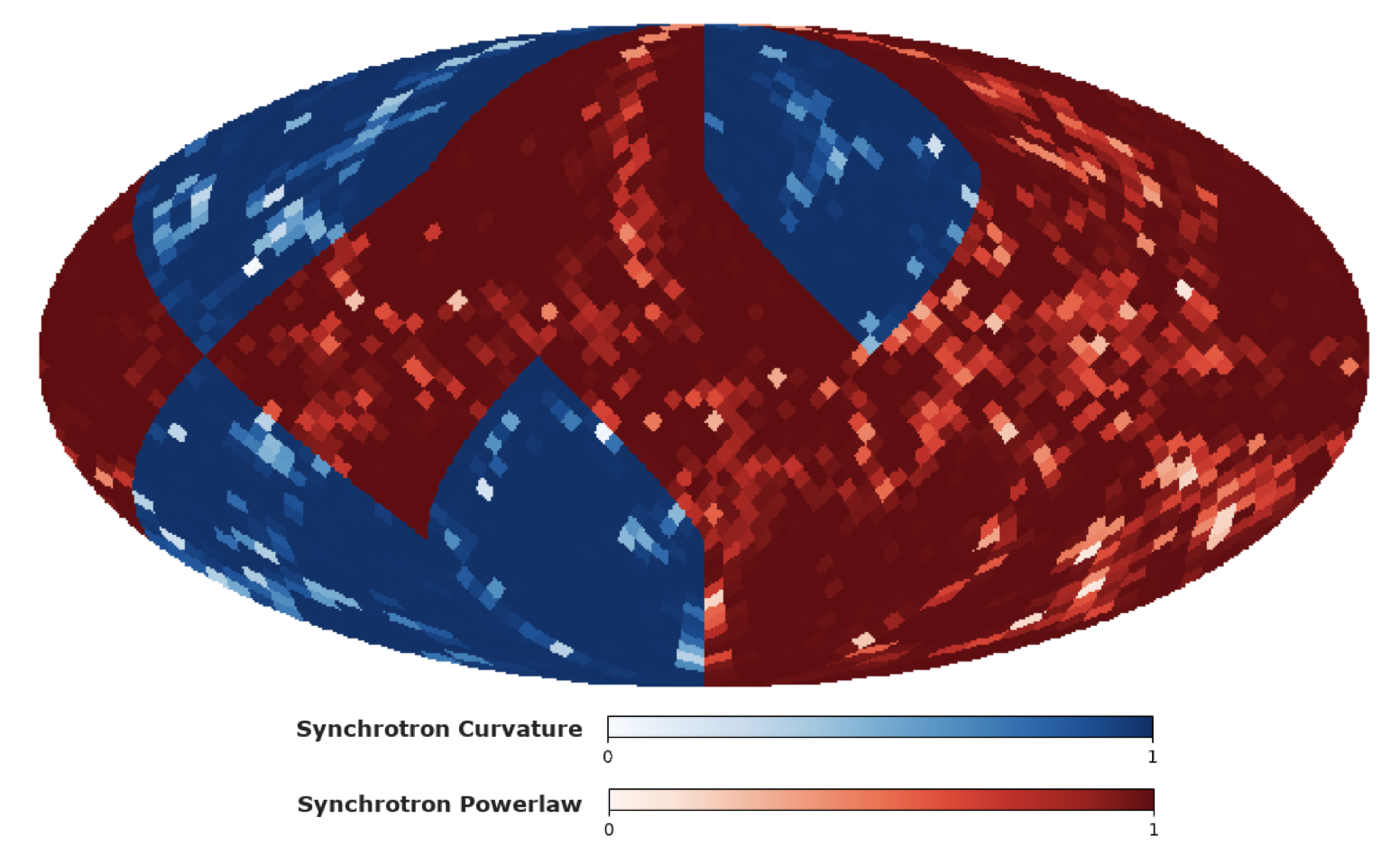}
\hspace{0.1cm}
\raisebox{0.88\height}{\includegraphics[width=0.3\textwidth]{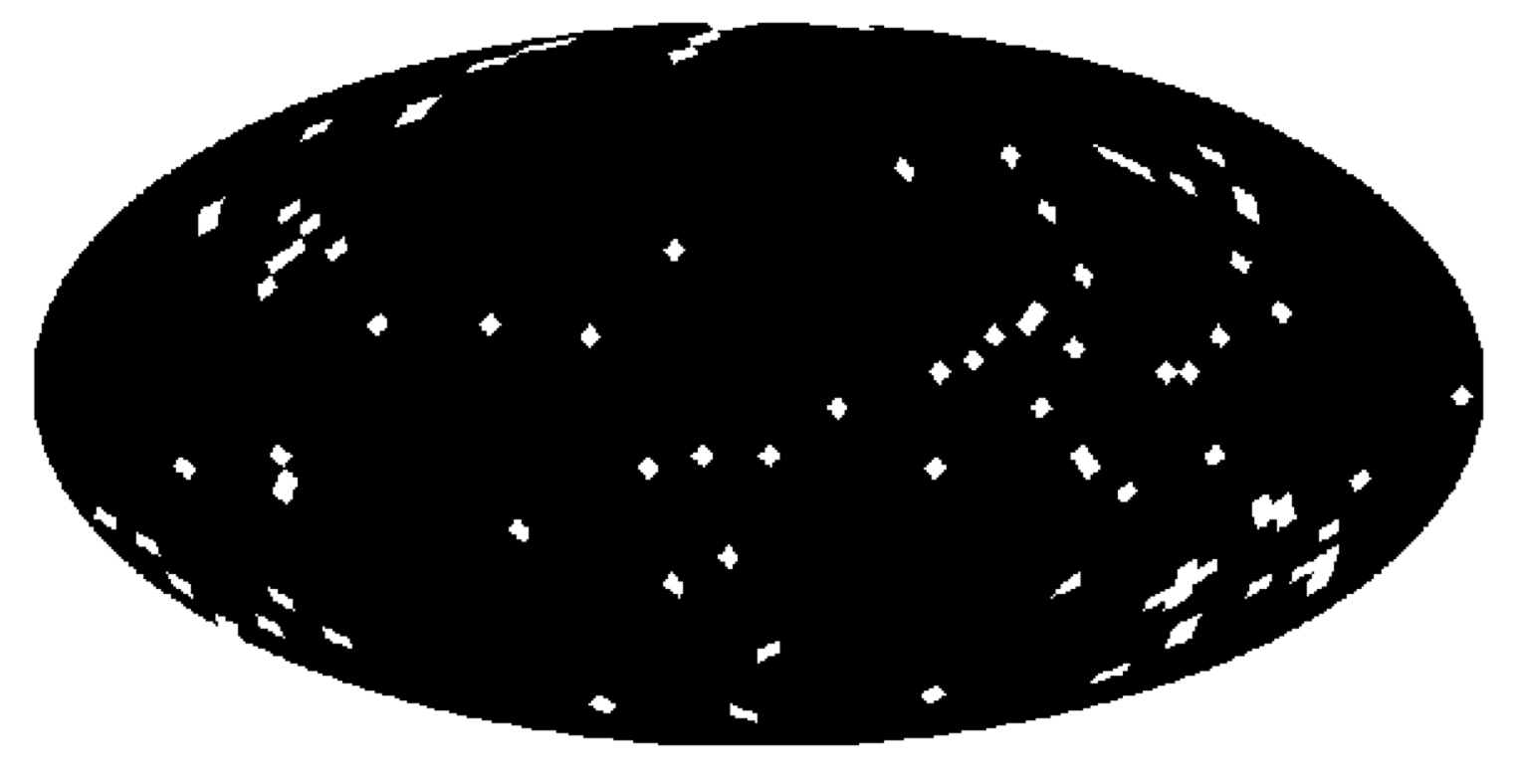}}
\caption{Comparison of $\chi^{2}$ analysis with NN prediction in the presence of noise on test maps. Upper panels: on the left, we report the probability for each pixel to belong to the correct model as obtained via the $\chi^2$ approach. In the red regions, the correct model is represented by a synchrotron power-law SED, while in the blue region a curvature is present. Lighter pixels are those where the $\chi^2$ analysis leads to a wrong model classification (also shown in white in the right panel). Lower panels: same as the upper panels but in this case the probability has been obtained via the NN approach. This comparison shows the advantage of using a NN approach, leading to a correct classification on about 97\% of the pixels with respect to about 73\% when the $\chi^2$ information is used.}
\label{fig:chi_square_S1S3}
\end{figure}

From the reduced $\chi^2$ we compute the probability for each pixel to belong to the correct model that we show in the upper panel of Figure \ref{fig:chi_square_S1S3}. As usual, darker colors indicate the pixels where thanks to the $\chi^2$ computation we retrieve the correct model, while lighter colors are for those pixels where the classification is wrong. We compare the results obtained from the $\chi^{2}$ with those of the NN (lower panel of \ref{fig:chi_square_S1S3}, in the case where we have re-trained the NN with 100 noise realization at $N_{side}=16$ (see Section \ref{sec:MC}). The reached accuracy calculated from reduced $\chi^{2}$ is at the level about $73\%$, while the NN is able to distinguish two models with of $97\%$ accuracy. This clearly shows the gain in using a NN for model recognition.\par
\begin{figure}
\centering 
\includegraphics[width=.6\textwidth]{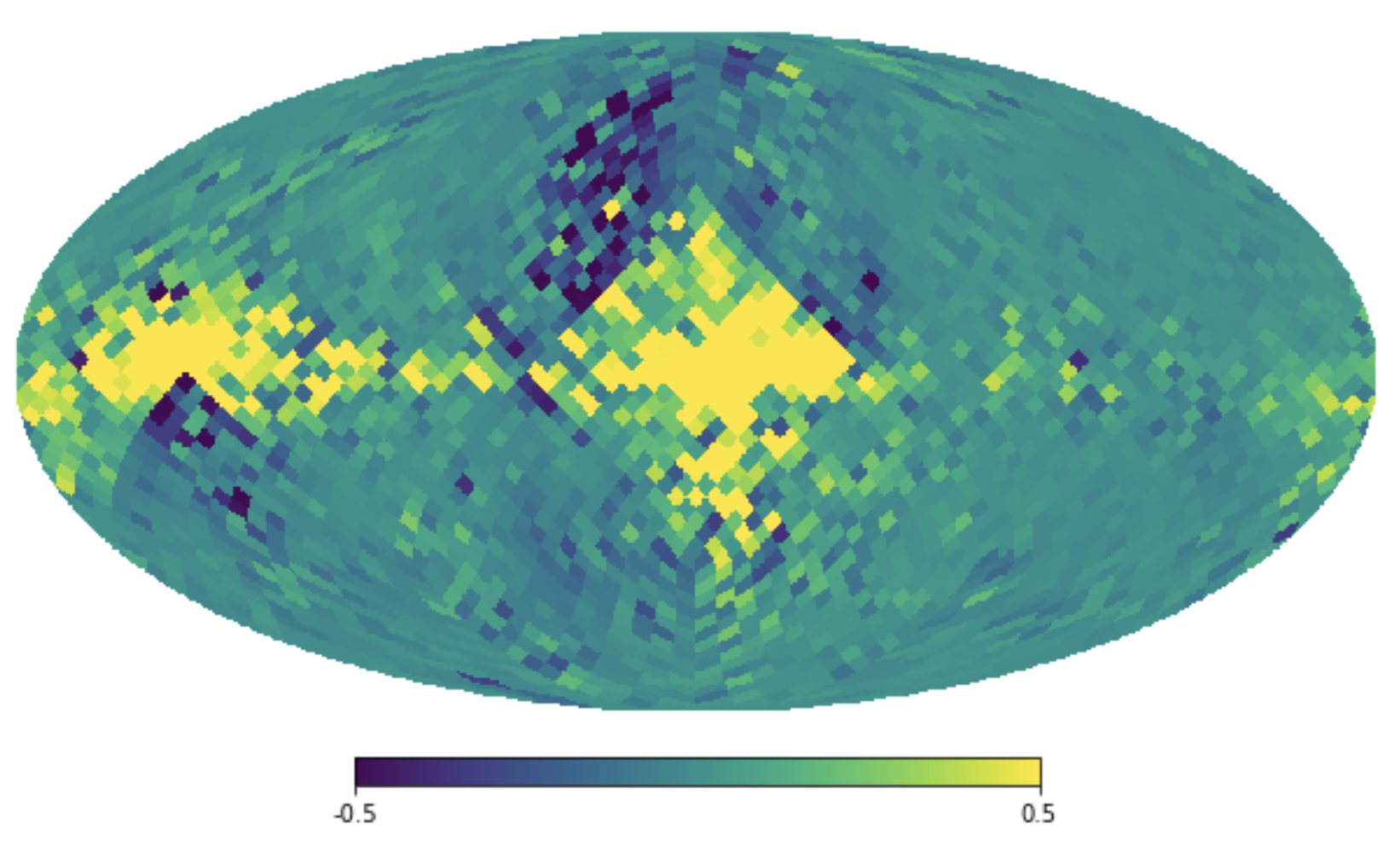}
\caption{Reduced $\chi^{2}$ difference for each pixel, obtained when the fit is done considering pure power-law SED for synchrotron and when the curvature is included. Pixels at intermediate and high Galactic latitudes (in green) are those where $\chi^2$ is unable to distinguish between the two models.}
\label{fig:chi_diff}
\end{figure}

In Figure \ref{fig:chi_diff} we also show the difference between the $\chi^{2}$ values computed in each pixel for the two different cases (with or without fitting for curvature) across the sky. As it is clear, the difference between the two reduced $\chi^{2}$ is very close to zero in the region where the sky signal is low (greenish regions at intermediate and high Galactic latitudes). These are the regions where the $\chi^{2}$ analysis leads to a higher probability of misclassification of the foreground model, due to the low signal-to-noise ratio. The same effect does not seem to affect the NN classification so strongly.  


\section{Conclusions}
\label{sec:conclusion}

In this work, we start to investigate the relevance of NN in recognizing the physical properties of the diffuse linearly polarized emission from our own Galaxy at microwave frequencies, which represents the main astrophysical contaminant to the measurement of the CMB $B$-mode polarization sourced by GWs in the early Universe. The problem is particularly challenging and urgent, due to scientific relevance of the cosmological signal, and the difficulty in disentangle it from the much brighter foreground emission.\par

Foreground cleaning is usually performed via parametric fitting, which implies the necessity of identifying the physical parameters describing the foreground model in each portion of the sky, fitting and marginalizing them on the basis of a suitable multi-frequency coverage. On the other hand, foreground physical properties and model do vary in the sky, in a manner which is currently only partially revealed by observations, and yet crucial, because the right parametrization of them is necessary to perform a good fitting and to prevent the presence of large foreground residual in the CMB maps which could bias the scientific results.\par 

In the present work, we study the possibility to identify the right physical parametrization of foregrounds, varying across the sky, in a pre-foreground cleaning phase. We do it with NNs, trained on simulations, and applied to test cases. We focus on the properties of Galactic synchrotron and AME, which have a rich phenomenology, resulting in possible different parametrization across the sky. We take care of making the simulations substantially different from observations, by explicitly and microscopically altering the training set with respect to the test one, at each resolution element. We find a good performance of the NN in recognizing the right parametrization of foregrounds, which achieve better results with respect to a standard $\chi^2$ test on the goodness of fit, making our results interesting and suitable for future studies.\par 

The combination of the simulations based on the specification of the QUIJOTE telescope and the LiteBIRD satellite, with a good coverage of the relevant frequencies, are analyzed in the binary and multi-class classifications modes, i.e. when two and four models have to be recognized in the sky, respectively. In all cases, the rate of success in recognizing the right foreground model is equal or larger than $90\%$. This is true even in the case where four foreground models have to be recognized, namely pure power-law SED with or without curvature for synchrotron, with and without AME. We compare the NN information concerning model recognition with the $\chi^{2}$ distribution following a parametric component separation assuming a given model, implemented and run through the publicly available {\tt FGBuster} code. We find that the NN perform better wit respect to the $\chi^{2}$, in particular at intermediate and high Galactic latitudes. 

We believe that these results are quite interesting, and a promising first step into the construction of a model recognition layer of data analysis in $B$-mode CMB measurements. Further lines of investigation concern the extension to other foreground models as well as the inclusion of possible realistic systematic effects. 

\acknowledgments
The authors thank Luca Heltai and Davide Poletti for useful discussions and suggestions. We acknowledge support from the ASI-COSMOS Network (cosmosnet.it) and by the INDARK INFN Initiative. 

\bibliographystyle{JHEP}
\bibliography{ref}

\end{document}